\documentclass[a4paper]{spie}  

\usepackage{amsmath,amsfonts,amssymb}
\usepackage{url}
\usepackage{graphicx}
\usepackage[colorlinks=true, allcolors=blue]{hyperref}
\usepackage{siunitx}
\usepackage{graphicx} 
\usepackage[numbers]{natbib}
\usepackage{aas_macros}

\usepackage{threeparttable}
\usepackage{diagbox}

\newcommand{\TT}{\texttt{TipTop}}
\newcommand{\maoppy}{\texttt{maoppy}}
\newcommand{\extraerr}{\texttt{extraErrorNm}}
\newcommand{\jitter}{\texttt{jitter\_FWHM}}

\title{Characterization of the MUSE NFM PSF as a function of atmospheric conditions: \TT\ calibration}

\author[a]{Enrico Congiu}
\author[a]{Fuyan Bian}
\author[b]{Carlo F. Manara}
\author[b]{Arseniy Kuznetsov}
\author[c]{Guido Agapito}
\author[d,e,f]{Johanna Hartke}
\author[e,d]{Timo Kravtsov}
\author[g]{Lisa-Marie Mazzolo}
\author[c]{Fabio Rossi}
\author[c]{Cedric Plantet}
\author[a]{Fernando Selman}

\affil[a]{European Southern Observatory (ESO), Alonso de Córdova 3107, Casilla 19, Santiago 19001, Chile;}
\affil[b]{European Southern Observatory (ESO), Karl-Schwarzschild Stra{\ss}e 2, D-85748 Garching bei M\"{u}nchen, Germany;}
\affil[c]{INAF -- Osservatorio Astrofisico di Arcetri, Largo E. Fermi 5, 50125, Firenze, Italy}
\affil[d]{Finnish Centre for Astronomy with ESO, (FINCA), University of Turku, 20014 Turku, Finland}
\affil[e]{Tuorla Observatory, Department of Physics and Astronomy, University of Turku, 20014 Turku, Finland}
\affil[f]{Turku Collegium for Science, Medicine and Technology (TCSMT), University of Turku, FI-20014 Turku, Finland}
\affil[g]{DOTA, ONERA, 13330, Salon-de-Provence, France}

\authorinfo{E.C.: E-mail: econgiu@eso.org}

\begin{document} 
\maketitle

\begin{abstract}
The Multi Unit Spectroscopic Explorer (MUSE) achieves exceptional spatial resolution in narrow-field mode (NFM) thanks to the GALACSI adaptive optics (AO) system. 
However, limitations in point spread function (PSF) characterization still hinder the full exploitation of its capabilities.
In particular, the current exposure time calculator (ETC) lacks an accurate PSF model, preventing users from reliably predicting the signal-to-noise ratio of NFM observations during proposal preparation.
To address this limitation, we analyzed a large set of archival standard-star observations to quantify how NFM PSF properties vary with observing conditions, including airmass, seeing, coherence time, wind speed, and wavelength.
We then used this reference dataset to calibrate TipTop, a fast AO PSF simulation tool that will be integrated into the next release of the MUSE NFM ETC.
Our results demonstrate that calibration against real on-sky data is essential for accurate PSF modeling. 
In particular, we find that reproducing realistic PSFs requires both an additional static aberration term and an airmass-dependent tip--tilt jitter component.
The calibration performs well at wavelengths longer than \qty{7000}{\angstrom}, while additional corrections are still required at shorter wavelengths, likely due to unmodeled chromatic aberrations.
Once implemented in the ETC, this tool will provide condition-dependent NFM PSF predictions and more reliable signal-to-noise estimates.

\end{abstract}


\section{Introduction}
\label{sec:intro}  

The exposure time calculator (ETC) is an essential tool for modern astronomical observatories, as it enables observers to carefully plan their observations and, as a consequence, observatories to allocate their available observing time efficiently.
Developing a reliable ETC requires several ingredients: accurate knowledge of the optical properties of the simulated system, a framework to simulate the observed targets, and a good understanding of how atmospheric conditions affect observations as a function of target position in the sky.
Among these, an accurate characterization of the point spread function (PSF), and in particular of its dependence on atmospheric turbulence, plays a central role.
The PSF governs how light from a point source is distributed across the detector, directly affecting the achievable signal-to-noise ratio (S/N) for a given exposure time.
Having a reliable model of the PSF is particularly important for facilities equipped with adaptive optics (AO) systems, where AO performance directly shapes it.
In these cases, simple analytical models such as Moffat or Gaussian profiles typically fail to describe the actual PSF, and the relationship between PSF properties, such as the full width at half maximum (FWHM) and Strehl ratio, and atmospheric conditions (seeing, wind speed, coherence time, etc.) is not straightforward.

In this context, the European Southern Observatory (ESO) is currently undertaking a comprehensive redevelopment of the ETCs for all its current and future instruments, with particular emphasis on incorporating accurate PSF models for AO-assisted facilities.
Among these, one of the first instruments to benefit from this effort will be the Multi Unit Spectroscopic Explorer (MUSE) integral field spectrograph, one of the most requested instruments at Paranal Observatory.
In its Narrow-Field Mode (NFM), MUSE operates with the GALACSI AO facility, using four Laser Guide Stars (LGSs) and one Natural Guide Star (NGS) to correct for atmospheric turbulence using a Laser Tomography AO (LTAO) technique.
This allows the system to deliver fully AO-corrected data over a \qtyproduct{7.5 x 7.5}{\arcsecond} field of view (FOV) and a $\sim$\qty{4500}{\angstrom} wavelength range in the optical band.
Under optimal conditions, GALACSI should enable MUSE to achieve near-diffraction-limited PSFs redward of 6000~\unit{\angstrom}, with Strehl ratios increasing toward longer wavelengths.
However, because of the spatial sampling of the NFM ($\sim$25~\unit{mas.per.pixel}) and the optical properties of the instrument, the best achievable FWHM is of the order of \qty{50}{mas}, while the Strehl ratio can be up to  $\sim$25\% at 9000~\unit{\angstrom} \cite{MUSEUserManual, Wevers22}.

In its current version, the MUSE ETC uses a fixed PSF model characterized by a wavelength-dependent FWHM, but independent from parameters such as target airmass and atmospheric conditions (seeing, coherence time).
This introduces uncertainties in the final S/N estimates, as well as in the expected spatial resolution under specific observing conditions.
To address this limitation, MUSE ETC v2.0 will introduce dynamic PSF simulations using \TT\ to generate a realistic MUSE NFM PSF for each ETC run.
\TT\ is an innovative tool that simulates AO-corrected PSFs from detailed information about the system and the specific observations \cite{TipTop}. 
Its integration, however, requires the definition of dedicated configuration files and its calibration against the actual system performance.

In this work, we describe the first steps in calibrating \TT\ for its integration into the MUSE ETC.
In Sec.~\ref{sec:tc} we briefly describe the ETC inputs that can affect the PSF.
In Sec.~\ref{sec:std} we describe how we compiled a database of comparison PSFs.
In Sec.~\ref{sec:tiptop} we present the calibration of \TT.
Finally, in Sec.~\ref{sec:summary} we summarize our conclusions.

\section{ETC inputs}
\label{sec:tc}

In order to reliably recover realistic S/N ratios for simulated observations, the ETC uses a wide variety of parameters.
However, only a relatively small subset affects the properties of the expected PSF: the turbulence category (TC), the airmass ($z$), the wavelength at which the simulations are performed, and the properties of the object used as NGS.
The TC combines into a single empirical parameter the two quantities most commonly used to characterize atmospheric turbulence in the context of AO systems: seeing (at zenith) and coherence time.
In practice, the TC describes the probability of acquiring an observation within the limits for these parameters reported in Tab.~\ref{tab:tc}.
In all categories, the seeing and the coherence time are measured by the DIMM \cite{Sarazin90} and MASS \cite{Kornilov03} systems at the Paranal Observatory.

\begin{table}[]
    \centering
    \caption{Definition of ESO turbulence categories. Each category is defined only if both conditions are satisfied at the same time.}
    \label{tab:tc}
    \begin{tabular}{lcc}
    \hline\hline
    TC & Seeing & Coherence time\\
        & (arcsec) & (ms)\\
    \hline
    10\%& $<0.60$ & $>5.2$\\
    20\%& $<0.70$ & $>4.4$\\
    30\%& $<0.80$ & $>4.1$\\
    50\%& $<1.00$ & $>3.2$\\
    70\%& $<1.15$ & $>2.2$\\
    85\%& $<1.40$ & $>1.6$\\
    \hline
    \end{tabular}
\end{table}

For MUSE NFM, the instrument is offered for science observations only in the first four categories, i.e., TC 50\% or better.
Calibration observations, such as standard stars, are sometimes acquired under slightly worse conditions.
In this work, we therefore define an additional category, TC 100\%, which includes all observations in our sample that do not satisfy the TC 50\% requirements.
While the TC system provides a reasonable characterization of atmospheric stability during observations, it nevertheless remains an oversimplification of the true atmospheric behavior.
In Sec.~\ref{sec:tiptop}, we discuss how this affects some of the assumptions required to calibrate \TT.

The airmass provides an estimate of the length of the atmospheric path that light must traverse before reaching the telescope.
Higher airmass values correspond to a longer path through the atmosphere and typically stronger atmospheric turbulence; as a consequence, AO performance generally degrades with increasing airmass.
Although MUSE NFM can observe up to an airmass of 2, corresponding to the minimum elevation allowed for laser propagation at the observatory, users are generally advised to observe targets below an airmass of 1.5, where the AO corrections are more reliable.
However, as for the TC, standard stars can occasionally be observed under less favorable conditions, and our analysis therefore includes observations acquired at airmass values up to 2.
It is also important to note that the TC system does not account for airmass: two targets observed under nominally identical TC conditions but at different airmass values can therefore exhibit significantly different PSFs.

It is well established that atmospheric turbulence varies with wavelength: the typical seeing decreases with increasing wavelength, following a power law with an exponent of $-1/5$ \cite{Fried66}.
This implies that the atmosphere is effectively more stable at longer wavelengths, making it easier for an AO system to correct for turbulence.
As a result, near-infrared (NIR) instruments can often achieve high Strehl ratios (50\% or more) and PSF widths comparable to the diffraction limit of the system.
The wavelength range covered by MUSE, however, is mostly optical and sufficiently broad that AO performance varies significantly across the spectrum, making it necessary to account for this variation in the ETC simulations.

Finally, GALACSI uses a near-infrared wavefront sensor called the InfraRed Low Order Sensor (IRLOS) to observe an NGS with a 2$\times$2 Shack--Hartmann sensor in the NIR (J+H-band) and correct low-order wavefront aberrations, such as focus and tip-tilt, that cannot be measured from the LGSs.
IRLOS operates in different modes depending on how much light it receives from the NGS and on whether the source is point-like or extended.
In particular, the frequency at which the sensor samples the wavefront decreases for faint and extended sources, reducing the AO correction performance.
For faint sources, the gain is also increased, boosting the noise and reducing the accuracy of the Shack--Hartmann spot centroid measurements.
Therefore, both the magnitude and the type of object used as NGS can significantly influence AO performance, as can its angular separation from the science target.
The most favorable configuration is when the NGS and science target are the same object (on-axis), while performance gradually degrades as the separation increases (off-axis), up to the physical limit of the instrument.

\section{Database of comparison PSF}
\label{sec:std}

To calibrate \TT\ so that it can produce realistic PSFs, we require a sample of well-characterized PSFs recovered from real observations obtained under a wide range of conditions.
Obtaining such a sample is not straightforward, as it requires observations of isolated point sources in different configurations and atmospheric conditions.
Moreover, a robust statistical analysis requires a relatively large number of observations.
Unfortunately, building such a sample from dedicated observations would be extremely expensive in terms of telescope time and would require repeated observations spanning several months, if not years.

We therefore decided to take advantage of standard star observations.
These are acquired every night the NFM is used in service mode, making it possible to build a significant sample from archival data.
In addition, the target is a single, typically isolated point source observed at relatively high S/N, making it well suited for estimating PSF properties.
Standard stars span a relatively wide range in J-band magnitude (8.8--\qty{14.9}{mag}), allowing us to sample a substantial fraction of the IRLOS dynamic range, and are acquired under a wide variety of atmospheric conditions, sometimes even outside the range typically offered to the community.

However, they are always observed on-axis, making it impossible to evaluate the effects of off-axis correction.
Astrometry fields, dense fields regularly observed to monitor the MUSE astrometric precision, could be useful in this context, especially when investigating the variation of the PSF in the field as a function of the distance from the AO star.
However, these calibrations are less frequent and the automatic on-mountain reduction does not produce datacubes for them, making them more difficult to exploit in an automated analysis pipeline.
In addition, even the astrometry fields cannot be used to investigate situations in which the AO star is not at the center of the field, as in that case the geometry of the NGS-LGS asterism will change, affecting the AO performances.
Standard stars also do not cover the brightest and faintest ends of the IRLOS dynamic range and are always point sources, preventing us from testing IRLOS performance on extended objects.
Nevertheless, most science observations fall within a parameter range that overlaps well with the standard star sample, and we therefore considered it a suitable reference for this initial analysis.

We compiled a catalog of standard star observations performed by MUSE between July 20, 2021 and March 23, 2026, totaling 302 data points.
We selected July 2021 as the starting point because the upgrade of the IRLOS sensor was completed on July 18, 2021, after which it began being used for science observations.
By limiting the analysis to observations acquired after this date, we ensure that the simulations are compared with data obtained using the same instrumental setup currently available at the telescope.

The data were processed with an analysis pipeline presented in detail in \cite{Wevers22} and originally developed by \cite{Hartke20} for MUSE-WFM data.
Minor updates were implemented for this work to improve the reliability of the PSF fitting for NFM data.
The pipeline has also been configured to run automatically every day on the Paranal Observatory infrastructure, providing PSF analysis for all MUSE observations.

In its NFM workflow, the pipeline uses \texttt{PampelMuse} \cite{Kamann2013} to fit a \maoppy\ \cite{Fetick2019} 2D profile to all sources identified and classified as stars by \texttt{SExtractor} \cite{Bertin96} in an I-band image extracted from the reduced MUSE cube.
The fit is performed across the full wavelength range.
All \maoppy\ parameters are free to vary except for $\alpha$ (the Moffat frequency transition parameter), which is fixed to 0.05.
This is necessary because the core of the MUSE NFM PSF is undersampled, and leaving $\alpha$ free can introduce degeneracies with other parameters such as $\rm r_0$.
A value of $\alpha=0.05$ has been shown by \cite{Fetick2019} and in the \maoppy\ documentation to be appropriate for this application.
At the end of this step, we obtain all the \maoppy\ parameters required to reconstruct the PSF in each wavelength channel.

Since \texttt{PampelMuse} does not directly return high-level quantities such as the Strehl ratio and FWHM, these must be determined a posteriori.
To do so, we extracted the average \maoppy\ parameters within $\pm$\qty{10}{\angstrom} of five representative wavelengths: 5000, 6000, 7000, 8000, and 9000~\unit{\angstrom}, and reconstructed the image of each PSF using the same \maoppy\ infrastructure.

We then measured the FWHM following an empirical approach.
First, we oversampled the PSF image by a factor of 10 to improve the measurement accuracy.
We then measured the distance from the peak at which the profile drops to half of its maximum value.
This computation was performed independently along the X and Y directions, and the average of the two measurements was adopted as the final FWHM.
To compute the Strehl ratio, we instead applied the \texttt{strehlOTF()} method included in the \maoppy\ package to the reconstructed models.

At this stage, we have a catalog of standard star observations whose PSFs have been characterized at five representative wavelengths across the MUSE spectral range, with FWHM and Strehl ratio measurements that can be used to compare with, and if necessary calibrate, \TT.

\section{TIPTOP calibration framework}
\label{sec:tiptop}

\begin{table}[]
    \centering
    \begin{threeparttable}
    \caption{Parameters of the \TT\ configuration file that depend on the specific observation. The first column represents the section of the configuration file they belong to, the second column the parameter name, and the third column the value assumed when the parameter is fixed.}
    \label{tab:TTparams}
    \begin{tabular}{lll}
    \hline\hline
    Section & Name & Fixed Value \\
    \hline
    \texttt{telescope}  & \texttt{ZenithAngle}   &    \\
    \texttt{telescope}  & \texttt{extraErrorNm}  &    \\
    \texttt{telescope}  & \texttt{jitter\_FWHM}  &    \\
    \texttt{atmosphere} & \texttt{Seeing}        &    \\
    \texttt{atmosphere} & \texttt{L0}            &    \\
    \texttt{atmosphere} & \texttt{Cn2Weights}    & $[$GLF, 1$-$GLF$]^*$ \\
    \texttt{atmosphere} & \texttt{Cn2Heights}    & $[$30m$^{*}$, 2000.0m$]$\\
    \texttt{atmosphere} & \texttt{WindSpeed}     &    \\
    \texttt{atmosphere} & \texttt{WindDirection} & $[$0, 0$]$\\
    \texttt{sensor\_HO} & \texttt{NumberPhotons} & $[$1740, 1740, 1740, 1740$]$ \\
    \texttt{source\_HO} & \texttt{Height} & 90000m*$z$ \\
    \texttt{sensor\_LO} & \texttt{Gain}          & \\
    \texttt{sensor\_LO} & \texttt{NumberPhotons} &  \\
    \hline\hline
    \end{tabular}
    \begin{tablenotes}
        \footnotesize
        \item[*] We assume \qty{30}{\meter} as the height of the ground layer, as this approximately corresponds to the height of the VLT domes.
    \end{tablenotes}
    \end{threeparttable}
\end{table}

\subsection{Observation-dependent parameters}

To generate a PSF with \TT, a configuration file must be defined, as described in detail in the \TT\ documentation \citep{TipTopDoc}. 
We divide the parameters into two categories: fixed parameters and observation-dependent parameters. Fixed parameters depend only on the system properties and remain unchanged once defined. 
Examples include the telescope diameter, the size of the central obstruction, and the number of elements in the Shack--Hartmann sensor. 
Observation-dependent parameters vary with the observing conditions and target properties. 
These include atmospheric quantities, such as seeing, wind speed and direction, and the turbulence profile, as well as observational quantities such as airmass and the flux measured by the wavefront sensors. 
Table~\ref{tab:TTparams} summarizes the parameters belonging to this second category.

For the parameters that do not depend directly on atmospheric conditions, we adopt different approaches depending on the quantity considered.
The \texttt{ZenithAngle} is computed from the observation airmass. The gain of the low-order (LO) sensor depends on the NGS magnitude. 
Following the values reported in the MUSE User Manual \citep{MUSEUserManual}, we assume a gain of 1 for stars brighter than 10.5~\unit{mag} in the J band, 68 for stars with magnitudes between 10.5 and 16~\unit{mag}, and 100 for fainter sources.

The remaining observation-dependent parameters are the \texttt{NumberPhotons} of the LO sensor (IRLOS) and the \texttt{Height} of the LGSs. 
To define the latter, we assume that the LGS altitude is primarily determined by the height of the sodium layer excited by the lasers. 
We adopt an average zenith height of \qty{90000}{\meter} and multiply it by the airmass to account for the increased distance of the sodium layer at lower elevations. 
It is important to note that the \TT\ treatment of the sodium layer does not include its vertical profile (e.g., thickness and density distribution) or its short-term variability. 
As a result, an important contribution to the real AO error budget is neglected, likely contributing to the calibration described in Sec.~\ref{sec:results}.

\begin{figure}
    \centering
    \includegraphics[width=0.7\textwidth]{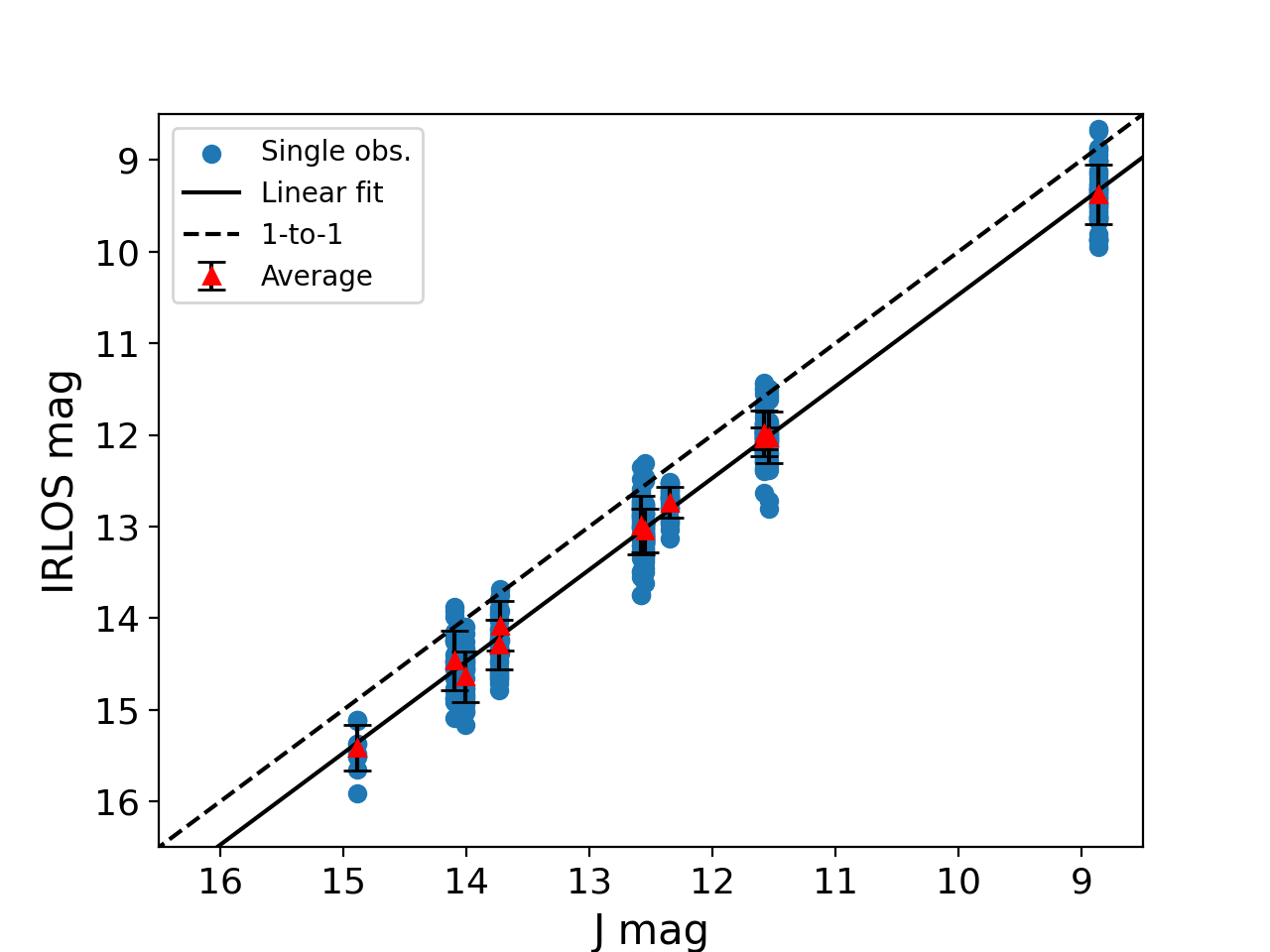}
    \caption{Relation between the J-band magnitude of the NGS and the magnitude observed by IRLOS.}
    \label{fig:ngs_mag}
\end{figure}

For the \texttt{NumberPhotons} parameter, we used our sample of standard stars to calibrate a relation between the J-band magnitude of the NGS and the magnitude measured by IRLOS, derived from the fluxes in ADU reported in the file headers. 
The data follow a linear relation with unit slope and a zero-point offset of 0.4~\unit{mag}. On average, IRLOS therefore measures stars as 0.4~\unit{mag} fainter than their cataloged J-band magnitudes, after correcting for airmass and atmospheric extinction in the J band (Fig.~\ref{fig:ngs_mag}).

We then use the relation in Eq.~\ref{eq:mag_to_flux} to convert the IRLOS magnitudes into the number of photons required by \TT.
In the equation, $J_0$ is the zero point of the J band in the AB system, $\mathrm{mag_J}$ is the J-band magnitude observed by IRLOS, and $F$ is the frequency of the IRLOS loop.

\begin{equation}
    \label{eq:mag_to_flux}
    N_{ph} = \frac{J_0 \times 10^{-0.4 \mathrm{mag_J}}}{4\times368\times F}
\end{equation}

We attempted a similar analysis for the \texttt{NumberPhotons} parameter of the high-order (HO) sensor (i.e., the LGSs), but, as a first approximation, we did not find any significant correlation between their flux and airmass or other parameters, despite the considerable variability of the LGS fluxes.
As a result, we currently treat this parameter as fixed, assigning to each LGS the median flux measured across the full standard-star sample.

\subsection{Atmospheric model assumptions}

The atmospheric parameters required by \TT\ differ from those directly available to the user in the ETC, and establishing analytical relations between them is not straightforward.
We therefore adopted a simple empirical approach.
First, we associated each observation in our standard-star sample with the corresponding atmospheric parameters measured during the observations.
All atmospheric data were obtained from the ESO ASM database\footnote{\url{https://www.eso.org/sci/facilities/paranal/astroclimate/ASMDatabase.html}}, while $L_0$ was extracted directly from the AO data logs acquired during the observations.
We then grouped the parameters by TC and analyzed their distributions.
Most quantities could be represented reasonably well by their median value, while others required additional assumptions.

\TT\ requires the turbulence ($C_n^2$) profile and the wind speed and direction arrays to contain the same number of entries.
However, wind at Paranal is sampled only at 10 and \qty{30}{\meter} above the ground.
Since \qty{30}{\meter} approximately corresponds to the height of the VLT domes, we assumed the wind properties measured at this level to be representative of the wind speed and direction throughout the atmospheric column.
In addition, the wind direction did not show a peaked distribution for which a meaningful average or median could be defined.
We therefore assumed a wind direction of \qty{0}{deg}, corresponding to wind coming from the north, the dominant wind direction at Paranal.

We also simplified the $C_n^2$ profile into a ground-layer component located at \qty{30}{\meter} and containing the fraction of turbulence below \qty{2000}{\meter}, and a high-altitude component located at \qty{2000}{\meter} containing the remaining turbulence.
Throughout the remainder of this work, we refer to the fraction of the $C_n^2$ profile located below \qty{2000}{\meter} as the ground layer fraction (GLF).
The final values adopted for each atmospheric parameter in the different TCs are presented in Tab.~\ref{tab:TCparams}.

\begin{table}[]
    \centering
    \caption{Values of the observation-dependent parameters as a function of turbulence category.}
    \label{tab:TCparams}
    \begin{tabular}{lccccc}
    \hline\hline
    Parameter & TC10\%& TC20\%& TC30\%& TC50\%& TC100\%\\
    \hline
    GLF & 0.970 & 0.973 & 0.980 & 0.972 & 0.977\\
    Wind Speed (\unit{\meter\per\second}) & 4.3 & 4.6 & 5.2& 5.9& 7.2\\
    L0 (\unit{\meter}) & 12.45 & 17.1 & 22.1 & 23.5 & 25.5\\
    \hline\hline
    \end{tabular}
\end{table}

\subsection{Calibration strategy}

With this setup, we are able to assign a value to each observation-dependent parameter listed in Tab.~\ref{tab:TTparams} for every simulation.
Early tests, however, showed that this setup is not sufficient for \TT\ to return realistic PSFs.
In particular, \TT\ tends to produce PSFs with much higher Strehl ratios than those observed in real data and with a FWHM of \qty{25}{mas}, exactly matching the MUSE NFM spaxel size.
This indicates that the simulated AO system performs better than the real one and that additional sources of error or aberrations must be introduced to better reproduce the actual system performance.

This can be done by acting on two parameters, \jitter\ and \extraerr, which are part of the \texttt{telescope} section.
Specifically, \jitter\ is the FWHM of a Gaussian convolution kernel applied to the PSF and represents an additional uncorrected source of tip--tilt jitter while \extraerr\ defines an additional Power Spectral Density (PSD) component added to the AO PSD to simulate generic aberrations, while (see the \TT\ documentation \cite{TipTopDoc} for details).
An additional parameter, \texttt{zCoefStaticOn}, allows the user to add static aberrations by activating specific Zernike modes when such aberrations are known to exist.
As this parameter is somewhat degenerate with the other two, we decided to limit the calibration to \extraerr\ and \jitter, since these primarily affect the Strehl ratio and the FWHM of the PSF, respectively, to first order.

The values of these parameters are not known \emph{a priori} and therefore need to be calibrated.
We do so by building samples of \TT\ PSFs obtained by randomly extracting observing conditions from uniform distributions in airmass, turbulence category, and NGS magnitude, and varying the values of \extraerr\ and \jitter\ for each sample.
We then compare these samples with the reference PSFs described in Sec.~\ref{sec:std} and identify the parameter values that best reproduce the observed PSFs.
The results of this optimization are presented in Sec.~\ref{sec:results}.

\section{Results and discussion}
\label{sec:results}

\subsection{Calibration at \qty{7000}{\angstrom}}

\begin{figure}
    \centering
    \includegraphics[width=0.9\textwidth]{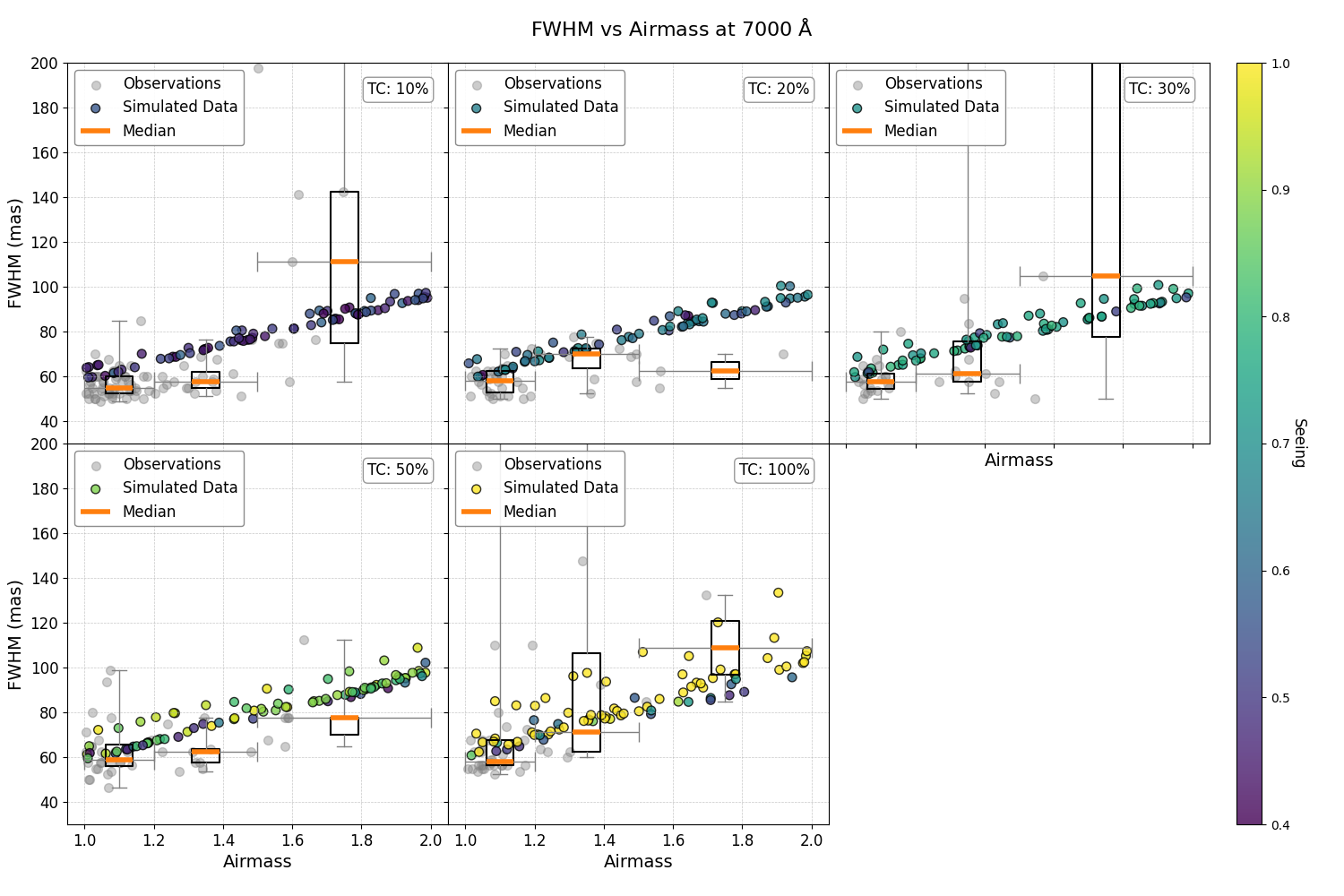}
    \caption{Observed and simulated FWHM at \qty{7000}{\angstrom} as a function of airmass and TC. Grey points represent the observations, while colored points represent the PSFs simulated by \TT. The color bar shows the value of the seeing assigned to the simulated observations. The orange bars show the median FWHM value in the airmass bins defined in Sec.~\ref{sec:std} for the observed PSFs. The associated bars show the 16th and 84th percentiles, while the vertical error bars indicate the full parameter range covered within each bin. The horizontal error bars show the width of the bins.}
    \label{fig:FWHM700}
\end{figure}

\begin{figure}
    \centering
    \includegraphics[width=0.9\textwidth]{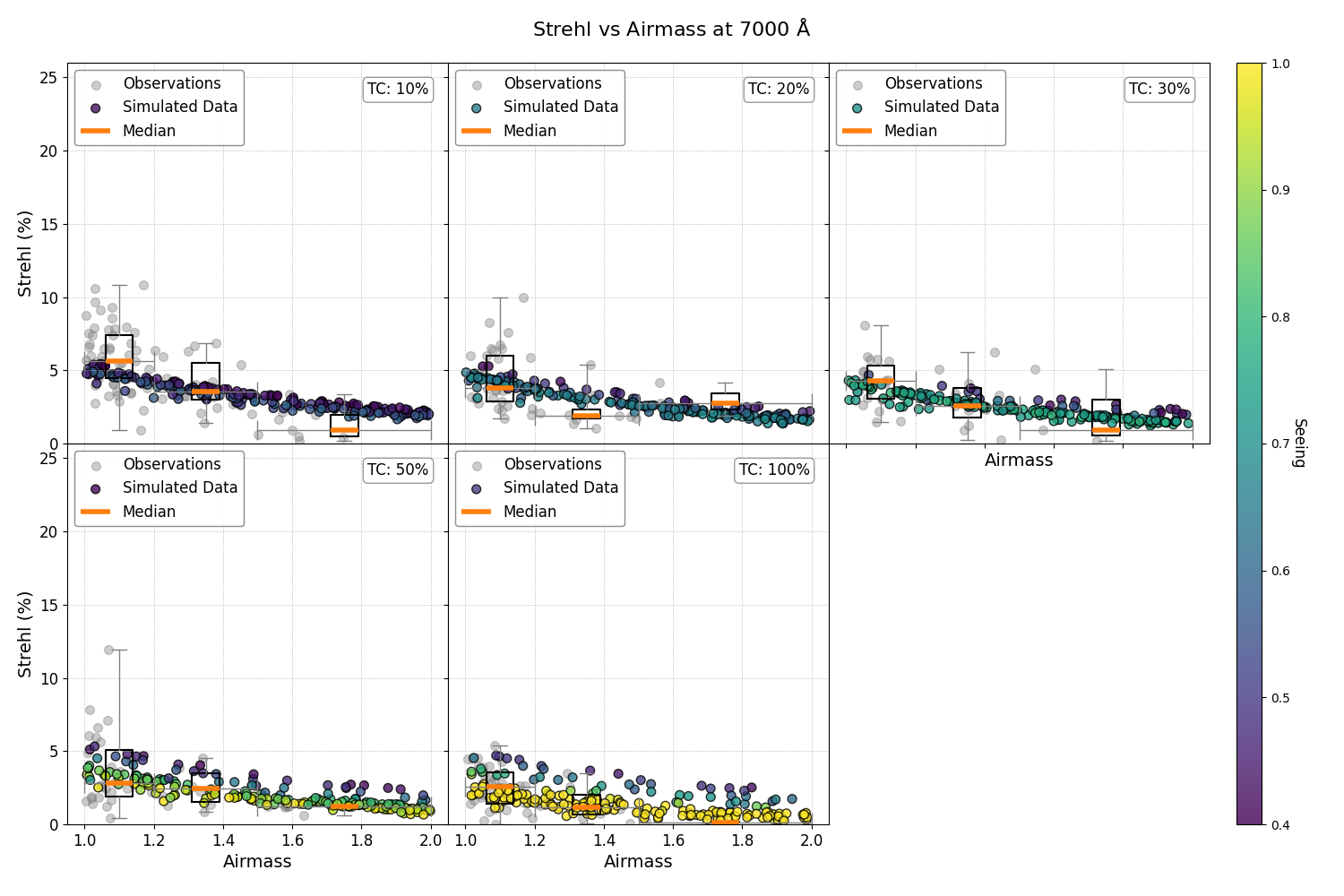}
    \caption{Observed and simulated Strehl ratios at \qty{7000}{\angstrom} as a function of airmass and TC. Symbols are the same as in Fig.~\ref{fig:FWHM700}.}
    \label{fig:STR700}
\end{figure}

\begin{table}[]
    \centering
    \caption{Optimal values for the \jitter\ and \extraerr\ parameters at \qty{7000}{\angstrom}. The \jitter\ is reported as the slope and intercept of the fit described in Sec.~\ref{sec:results}.}
    \label{tab:tt700}
    \begin{tabular}{lc}
    \hline\hline
    Parameter & Value \\
    \hline
    Slope & 28.7\\
    Intercept  & 25.9\\
    \extraerr  & 170\\
    \hline\hline
    \end{tabular}
\end{table}

In Sec.~\ref{sec:std} we characterized our standard-star sample at five different wavelengths (5000, 6000, 7000, 8000, and 9000~\unit{\angstrom}).
However, in the absence of chromatic aberrations, calibrating the PSF at a single wavelength should also produce realistic PSFs at the other wavelengths.
We therefore start our calibration at \qty{7000}{\angstrom} since this value is in the middle of the available MUSE wavelength range.

As the \extraerr\ and \jitter\ parameters have relatively independent effects on the Strehl ratio and FWHM of the final PSF, they can be optimized almost independently.
We started by optimizing the \jitter\ parameter.
We found that a constant \jitter\ value was unable to reproduce the airmass dependence of the FWHM observed in our reference sample.
This could indicate that AO performance degrades with increasing airmass in a way that is not fully accounted for in the current \TT\ simulations.
This may be because some of the assumptions adopted in Sec.~\ref{sec:tiptop} are too simplistic to fully represent the variety of observing conditions encountered at the observatory, or because some elevation-dependent aberrations are not fully taken into account.
One example could be the misregistration between the wavefront sensor and the deformable secondary mirror, caused by mechanical flexures and drifts that occur on short timescales during observations.
The AO facility automatically updates the control matrices to compensate for these flexures \cite{Oberti18} but between updates residuals transient errors can appear, affecting the AO performances.
As \TT\ relies on an average, static, representation of the system, these errors cannot be reproduced and they could contribute to the reason why the additional sources of errors need to be injected to reproduce the observed PSF.

To estimate the amount of \jitter\ required as a function of airmass, we fitted the FWHM--airmass relation and derived its slope and intercept.
For this fit we relied only on the observations at $z<1.5$, since they are those for which the AO performances can be considered mostly reliable.
We then estimated the \jitter\ value required at each airmass by computing the expected FWHM from the fit and subtracting in quadrature the FWHM returned by \TT\ (\qty{25}{mas}) in absence of additional error terms.
This is equivalent to assuming a Gaussian PSF core and deriving the FWHM of the Gaussian convolution kernel required to reproduce the expected PSF.
The FWHM--airmass relation at \qty{7000}{\angstrom} does not appear to depend strongly on TC; therefore, we performed a unique fit for all TCs.
The slope and intercept of the relation are reported in Tab.~\ref{tab:tt700}.

To estimate the optimal \extraerr\ value, we started from the PSFs obtained using the \jitter\ values derived above and progressively increased \extraerr\ from zero in steps of 10 units until the distribution of simulated Strehl ratios overlapped with that of the reference sample.
The optimal \extraerr\ value is also reported in Tab.~\ref{tab:tt700}.

Figures~\ref{fig:FWHM700} and \ref{fig:STR700} show the results of this optimization.
As expected, the simulated and observed PSFs cover a very similar parameter space.
The observed PSFs show a significantly larger scatter, which is partially reproduced only for the worst TC.
The origin of this scatter is not clear.
The simulations suggest that in the worst TCs, the spread could be caused by the broad range of values that the seeing can assume.
However, this seems not to be the case for the best turbulence categories.
Another possibility is that the DIMM and MASS sample a specific line of sight, that often does not coincide with the line of sight of the telescope.
As such, conditions could vary and this would cause observations to be misclassified.
Finally, wind-shake and dome seeing depend heavily on the relative geometry between the wind vector and the dome opening, and both can vary significantly over short timescales.
This could be driving the scatter in the observed PSF, and as these effects are extremely difficult to model, \TT\ is not able to reproduce them.
Such effects could also be somewhat elevation dependent, and contribute to the airmass dependence of the measured \jitter.

Figure~\ref{fig:STR700} also shows that, particularly for good TC, the Strehl--airmass relation obtained from the simulations is flatter than the observed one, although it remains consistent with the error bars of the binned measurements.
Also in this case, the origin of this behavior is not clear, but it may be related to the same effects discussed above for the airmass dependence of the \jitter\ parameter.
A more in-depth analysis of the GALACSI AO system and its representation in \TT\ is needed to clarify these points.
In addition, a more refined definition of the observing conditions in the ETC, better representing the quantities relevant for AO-assisted observations, could help users obtain more realistic estimates of the PSF and S/N.

\subsection{Multiwavelength calibration}

\begin{table}[]
    \centering
    \caption{New values of the intercept for the FWHM--airmass relation at 5000 and \qty{6000}{\angstrom}.}
    \label{tab:ttblue}
    \begin{tabular}{lcc}
    \hline\hline
    Parameter & \qty{5000}{\angstrom} & \qty{6000}{\angstrom} \\
    \hline
    Intercept  & 15.9 & 20.9\\
    \hline\hline
    \end{tabular}
\end{table}

To validate the assumption that calibrating the PSF at a single wavelength is sufficient to reproduce the MUSE NFM PSF across the full wavelength range, we generated \TT\ PSFs at all the wavelengths examined in Sec.~\ref{sec:std} using the same \jitter\ and \extraerr\ values calibrated at \qty{7000}{\angstrom}.
The results are presented in Figs.~\ref{fig:FWHM500}--\ref{fig:STR900} in Appendix~\ref{sec:app}.

The parameters tuned at \qty{7000}{\angstrom} accurately reproduce the Strehl–airmass relationship across all wavelengths and for every TC.
In contrast, for the FWHM, although the simulated values consistently lie within the range measured in the reference sample, they systematically tend to be higher than observed at the shorter wavelengths.

At \qty{5000}{\angstrom}, in particular, we observe a significant offset of about $\sim$\qty{20}{mas} under the best observing conditions.
For worse conditions (TC$>$50\%), the offset instead becomes a large spread, partially reproducing the scatter observed in the reference sample.
At \qty{6000}{\angstrom}, the behavior is similar, although both the offset in good conditions ($\sim$\qty{10}{mas}) and the spread in poor conditions are smaller.
At longer wavelengths (8000 and \qty{9000}{\angstrom}), instead, the parameters calibrated at \qty{7000}{\angstrom} produce excellent agreement for both the FWHM and Strehl ratio.

This behavior suggests that chromatic aberrations are indeed present in the system and affect short wavelengths more strongly than long wavelengths.
The origin of these aberrations is currently unknown.
One possibility is that they are related to non-common-path aberrations in the instrument.
If produced by transmissive optical elements, such aberrations could introduce chromatic effects.
Unfortunately, a detailed characterization of non-common-path aberrations in MUSE has never been performed, and this is outside the scope of the present work.

For these two wavelengths, we repeated the calibration procedure while keeping fixed both the \extraerr\ value and the slope of the \jitter\ versus airmass relation, since the Strehl--airmass relation and the slope of the FWHM--airmass relation already appear to be well reproduced.
We therefore varied only the intercept of the relation, in steps of \qty{5}{mas}.
We adopted this approach instead of the one described in Sec.~\ref{sec:tiptop} because the large scatter at poor TC significantly biased the linear fit.
The new intercept values are reported in Tab.~\ref{tab:ttblue}, while Fig.~\ref{fig:FWHM500_2} and Fig.~\ref{fig:FWHM600_2} show the updated point distributions.

\section{Summary and Conclusions}
\label{sec:summary}

In this work, we compiled a sample of PSFs from MUSE NFM observations of standard stars and used it, together with an analysis of the paranal atmospheric conditions, to calibrate the \TT\ parameters required to predict the MUSE PSF as a function of observing conditions and airmass. 
As the final goal was to integrate the results of this analysis into the new version of the MUSE ETC, which will soon be released, the analysis was performed using only parameters typically available to the user in this context, namely TC, airmass, and NGS magnitude. 
The main calibration was performed at \qty{7000}{\angstrom}, which is the reference wavelength used at the observatory to characterize the MUSE PSF, but it was also extended to both shorter and longer wavelengths.

With this analysis, we found that there is a significant spread in the FWHM and Strehl ratios measured for sources observed under very similar conditions. 
The origin of this scatter is not clear, but it could be related to the TC definition, and in particular to the fact that the atmospheric conditions sampled by the DIMM and MASS may not be fully representative of those along the instrument line of sight during the observations. 
In addition, factors such as dome seeing and wind-shake could also play a significant role.

We then found that, when relying only on our basic assumptions for the observing setup and atmospheric conditions, \TT\ produces significantly better PSFs than those observed in our reference sample. 
Injecting additional errors through the \jitter\ and \extraerr\ parameters is therefore essential to reconcile the simulations with the observations. 
These additional error terms may have several origins: intrinsic non-common-path aberrations not included in the \TT\ model, an oversimplified representation of the atmospheric conditions, misregistration between the wavefront sensor and deformable mirror not accounted for in the simulations, and an oversimplification of \TT\ treatment of the atmospheric sodium layer.

A constant value of \extraerr\ allows \TT\ to reproduce the Strehl ratios observed at \qty{7000}{\angstrom} reasonably well, although the simulated Strehl--airmass relation obtained under good conditions remains flatter than the observed one. 
However, \jitter\ requires an airmass-dependent calibration to reproduce the observed FWHM distribution. 
This suggests that AO performance degrades with increasing airmass in a way that is not fully accounted for by \TT. 
This effect, as well as the discrepancy observed in the Strehl--airmass relation, could be caused either by an oversimplified description of the observing conditions or by additional elevation-dependent aberrations not included in the simulations, such as misregistration between the wavefront sensors and the deformable mirror.

Extending the analysis to other wavelengths, the \extraerr\ calibration obtained at \qty{7000}{\angstrom} works well across both the red and blue parts of the MUSE wavelength range. 
The \jitter\ calibration, on the other hand, performs well only at wavelengths redder than \qty{7000}{\angstrom} and requires additional corrections to reproduce the FWHM observed at \qty{5000}{\angstrom} and \qty{6000}{\angstrom}. 
This points toward the presence of chromatic aberrations within the instrument that are currently not characterized.

This study represents one of the first efforts to use \TT\ to predict the MUSE NFM PSF using only parameters available to the user through the ETC, and as such differs from other works, such as \citet{Kuznetsov26}, whose goal is to reconstruct the PSF starting from a detailed analysis of the conditions associated with a specific observation. 
Like any initial effort, however, it has several limitations. 
In particular, the reference sample only includes relatively bright on-axis NGSs and is therefore not fully representative of the full range of observations that can be performed with the instrument.
In addition, the characterization of the atmospheric parameters remains relatively simplistic compared to the actual observing conditions at Paranal, particularly in the treatment of the turbulence profile. 
While the adoption of a simplified two-layer model based on the GLF proved sufficient for this initial calibration, it remains a known limitation when simulating LTAO systems. 
Because LTAO performance depends strongly on the vertical distribution of turbulence through tomographic reconstruction and focus anisoplanatism, collapsing all high-altitude turbulence into a single layer at \qty{2000}{\meter} inevitably limits the realism of the simulations. 
Extending the framework to support multi-layer turbulence profiles is therefore a natural next step and is expected to reduce the discrepancies between simulated and observed PSFs.
It may also be necessary to develop a more suitable scheme for describing observing conditions when such classifications are required for operational or user-support purposes.

Finally, even though MUSE and GALACSI have been operating jointly for nearly a decade, several elements of the system remain insufficiently characterized and could substantially enhance the accuracy of simulation tools like \TT. 
In particular, a detailed knowledge of the non-common-path aberrations of the system would allow them to be directly injected into the simulations and reduce the need to rely on empirical calibrations. 
Potential future upgrades also include a better understanding of the impact of off-axis, faint, and extended NGSs.

\acknowledgments 
JH and EC acknowledge the financial support from the Visitor and Mobility program of the Finnish Centre for Astronomy with ESO (FINCA).

\bibliography{report} 
\bibliographystyle{spiebib} 

\begin{appendix}
\section{Additional Figures}
\label{sec:app}

\begin{figure}[h!]
    \centering
    \includegraphics[width=0.9\textwidth]{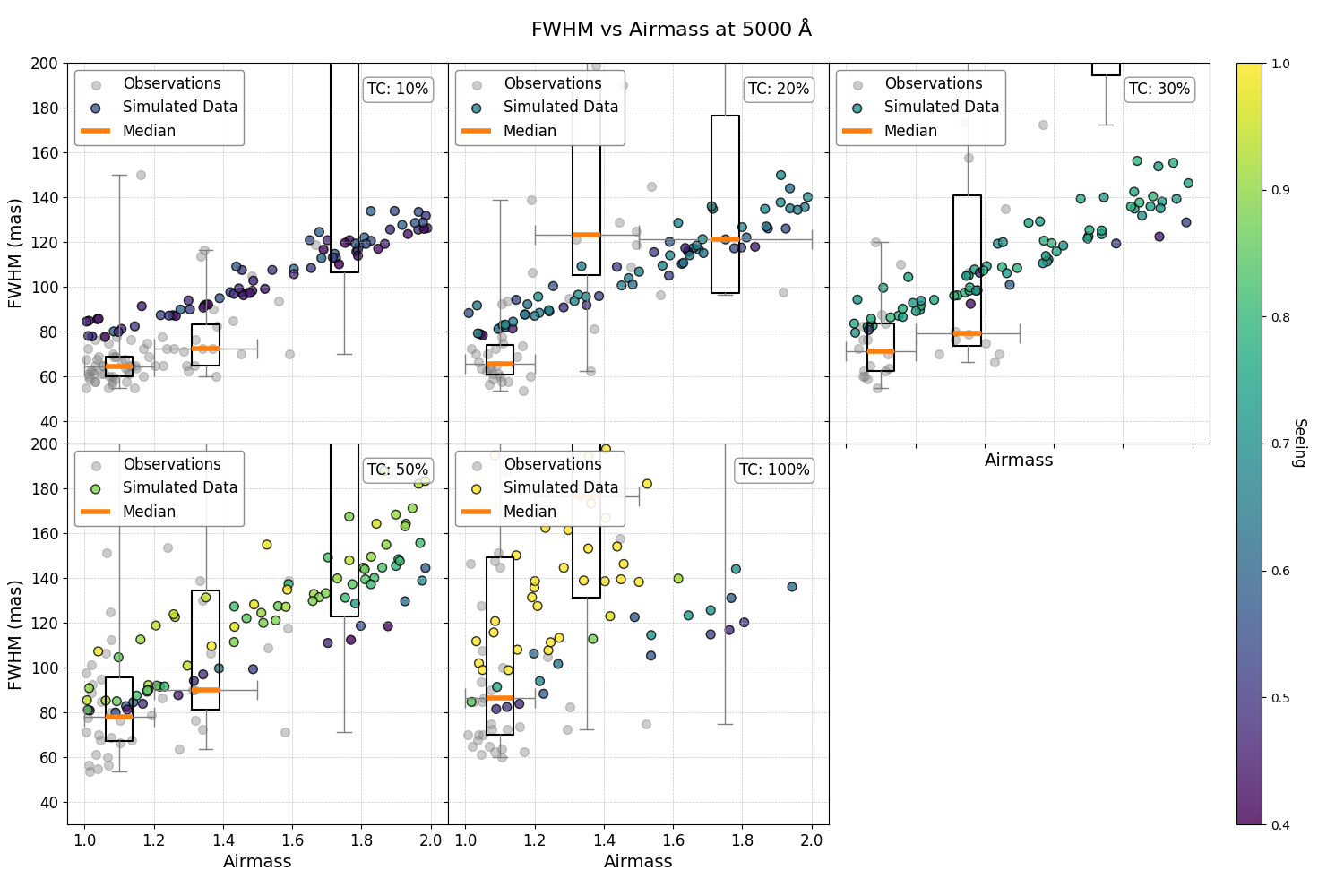}
    \caption{Observed and simulated FWHM as a function of airmass and TC at \qty{5000}{\angstrom}. Symbols as for Fig.~\ref{fig:FWHM700}}
    \label{fig:FWHM500}
\end{figure}

\begin{figure}
    \centering
    \includegraphics[width=0.9\textwidth]{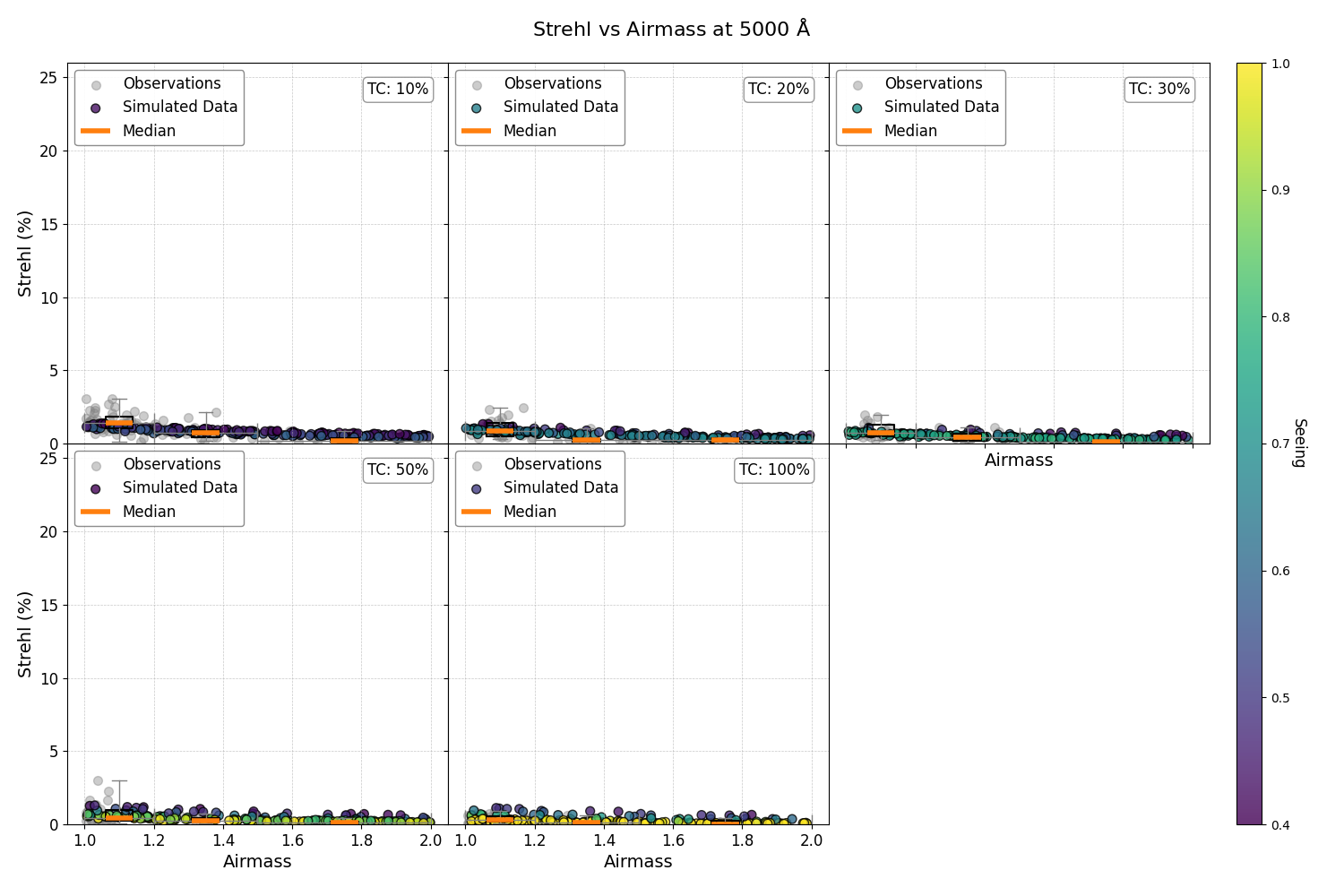}
    \caption{Observed and simulated Strehl ratios as a function of airmass and TC \qty{5000}{\angstrom}. Symbols as for Fig.~\ref{fig:FWHM700}}
    \label{fig:STR500}
\end{figure}

\begin{figure}
    \centering
    \includegraphics[width=0.9\textwidth]{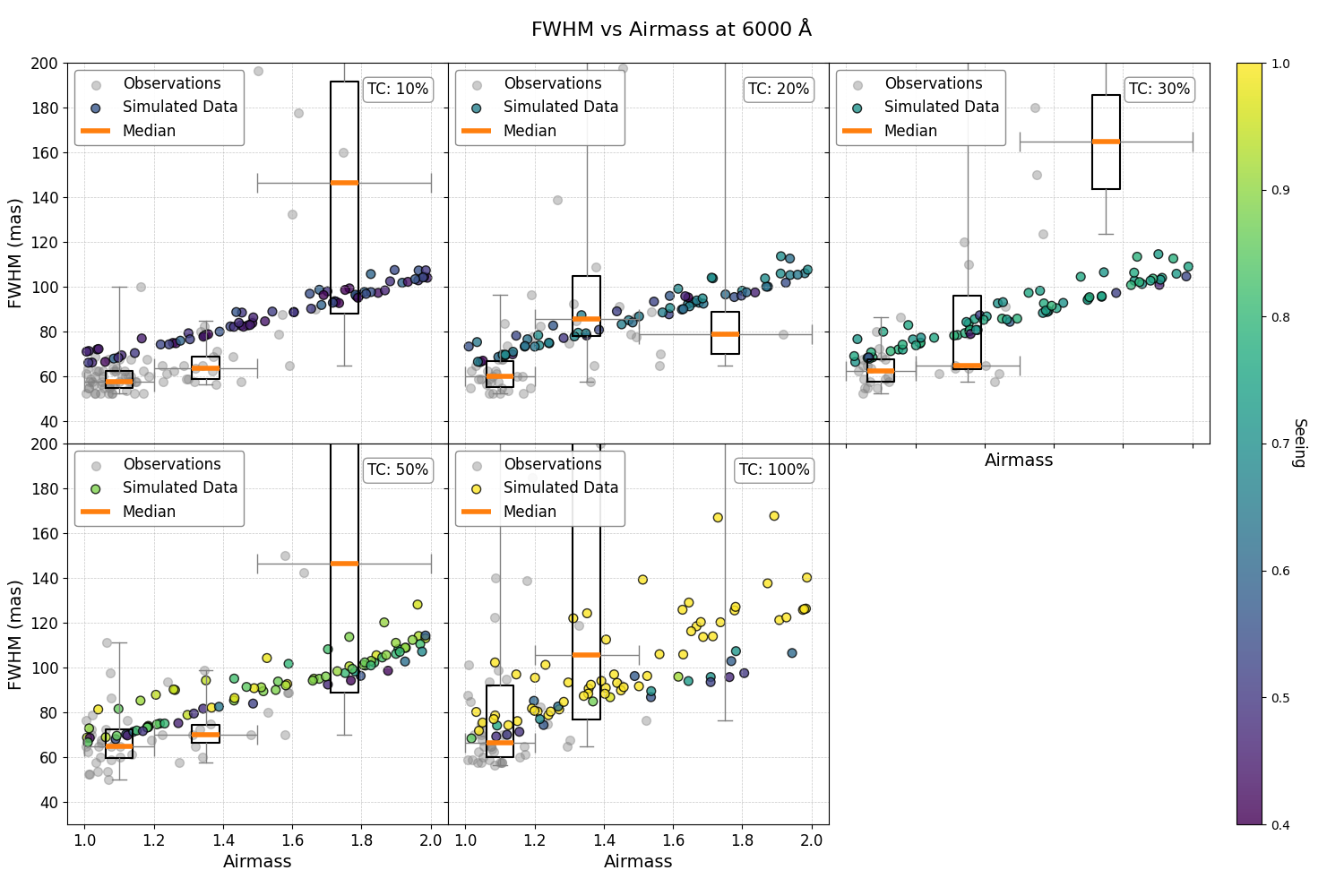}
    \caption{Observed and simulated FWHM as a function of airmass and TC at \qty{6000}{\angstrom}. Symbols as for Fig.~\ref{fig:FWHM700}}
    \label{fig:FWHM600}
\end{figure}

\begin{figure}
    \centering
    \includegraphics[width=0.9\textwidth]{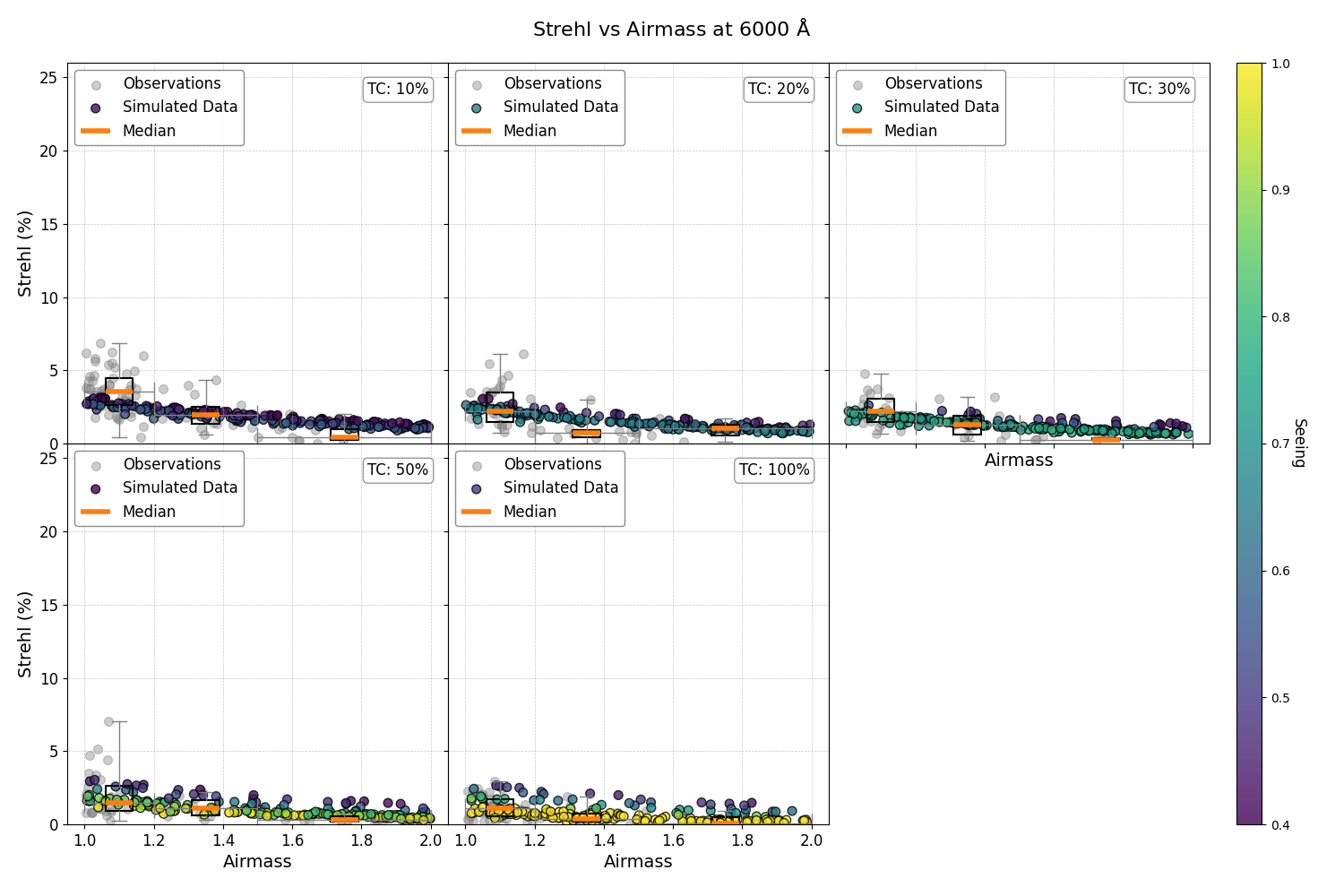}
    \caption{Observed and simulated Strehl ratios as a function of airmass and TC \qty{6000}{\angstrom}. Symbols as for Fig.~\ref{fig:FWHM700}}
    \label{fig:STR600}
\end{figure}

\begin{figure}
    \centering
    \includegraphics[width=0.9\textwidth]{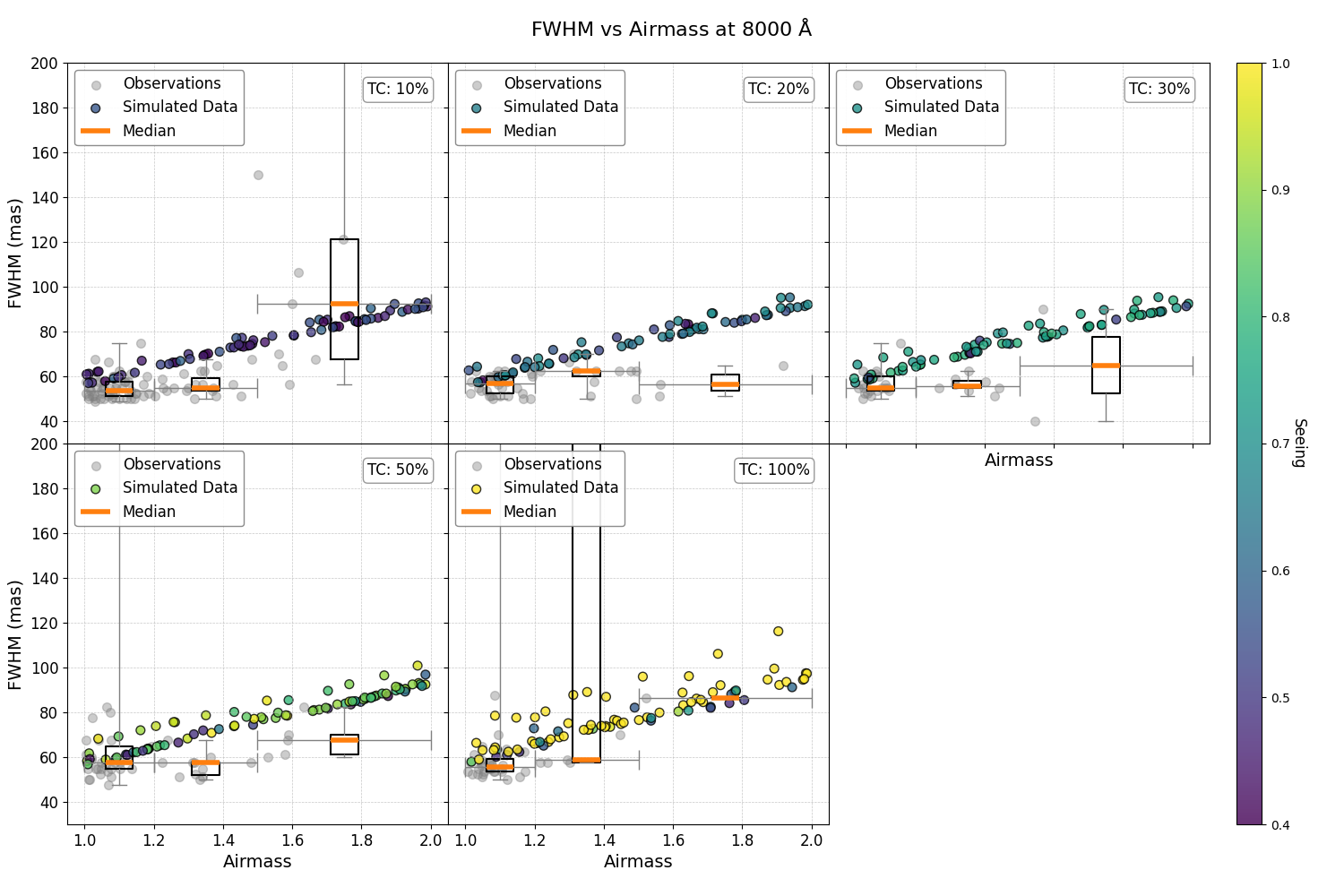}
    \caption{Observed and simulated FWHM as a function of airmass and TC at \qty{8000}{\angstrom}. Symbols as for Fig.~\ref{fig:FWHM700}}
    \label{fig:FWHM800}
\end{figure}

\begin{figure}
    \centering
    \includegraphics[width=0.9\textwidth]{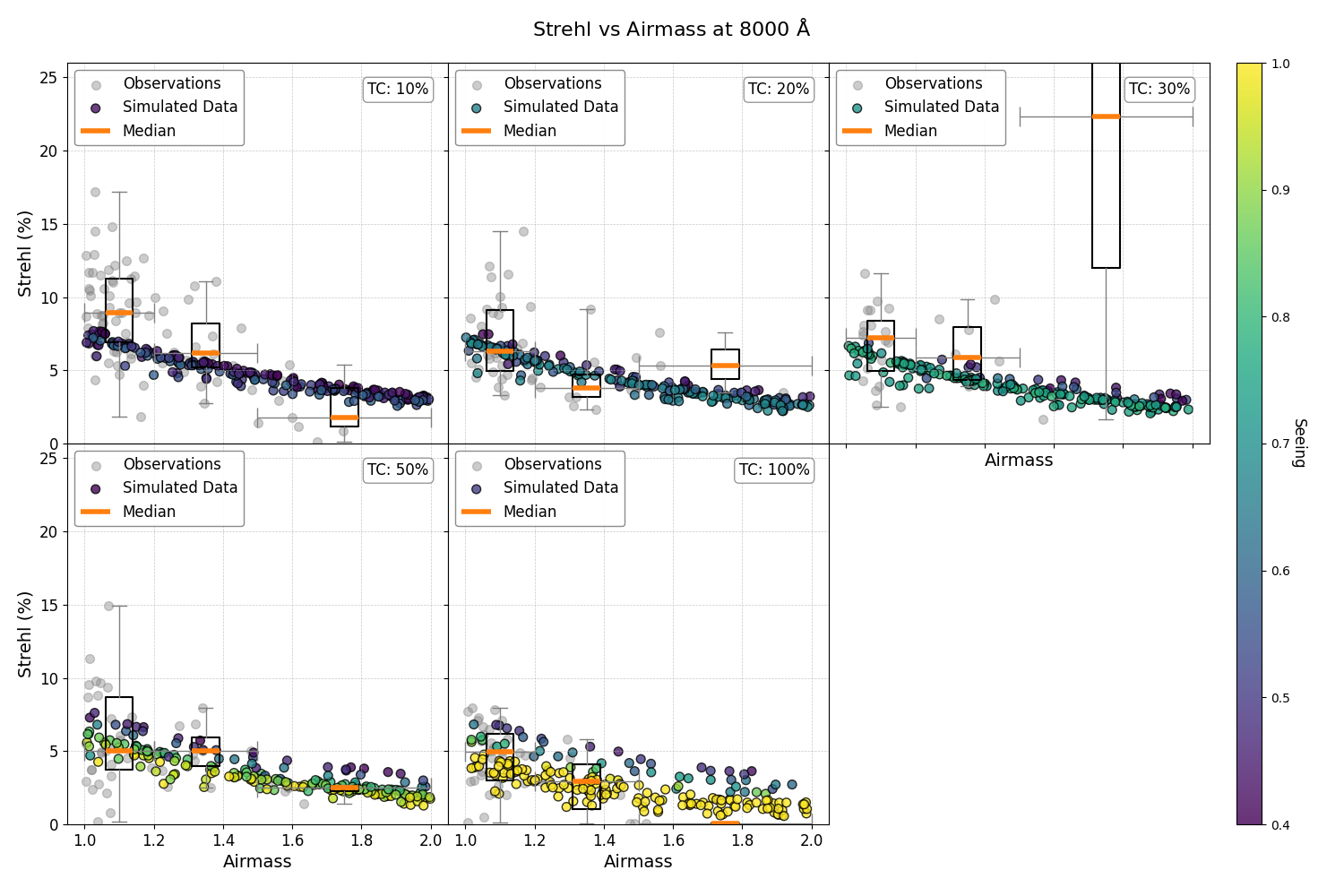}
    \caption{Observed and simulated Strehl ratios as a function of airmass and TC \qty{8000}{\angstrom}. Symbols as for Fig.~\ref{fig:FWHM700}}
    \label{fig:STR800}
\end{figure}

\begin{figure}
    \centering
    \includegraphics[width=0.9\textwidth]{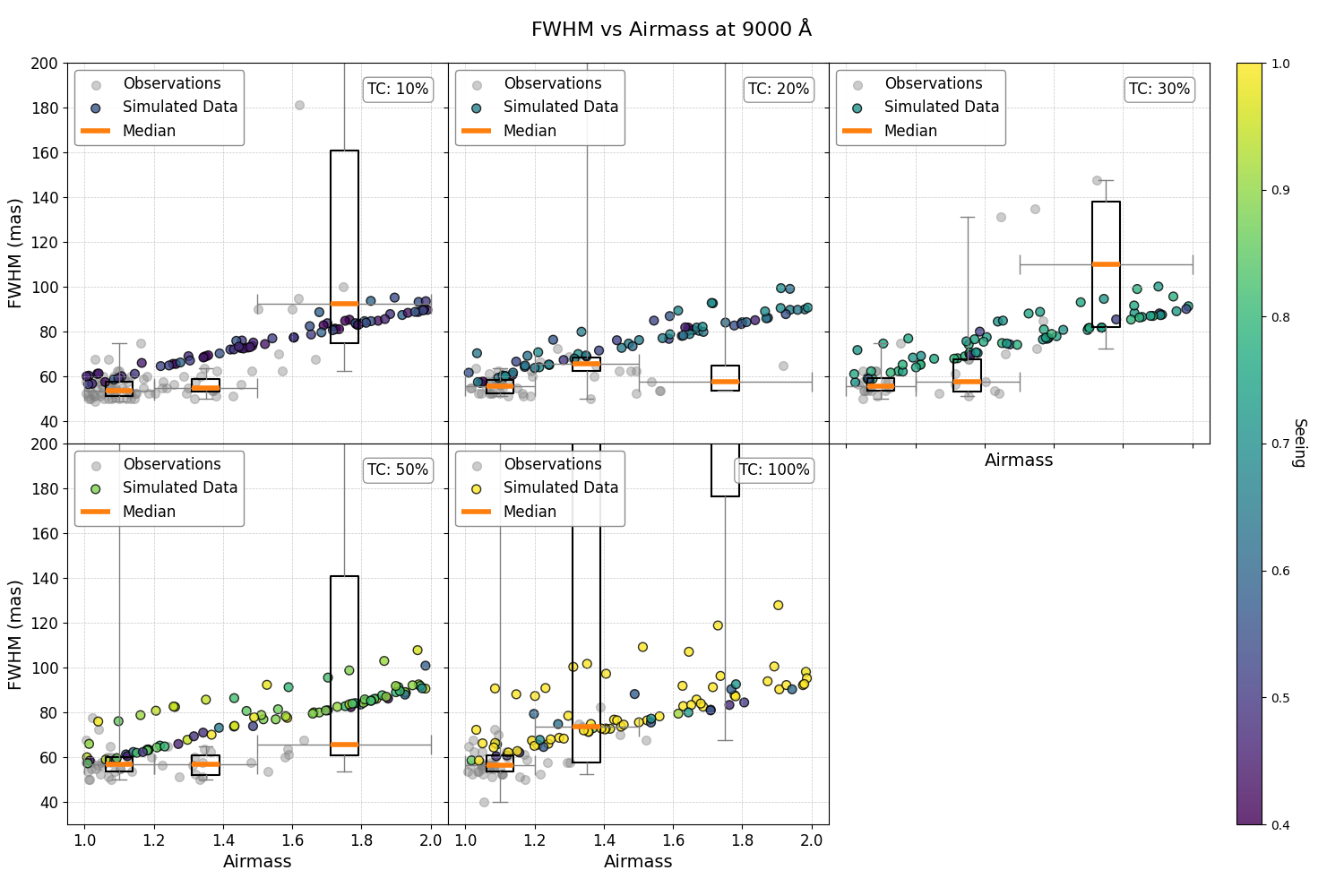}
    \caption{Observed and simulated FWHM as a function of airmass and TC at \qty{9000}{\angstrom}. Symbols as for Fig.~\ref{fig:FWHM700}}
    \label{fig:FWHM900}
\end{figure}

\begin{figure}
    \centering
    \includegraphics[width=0.9\textwidth]{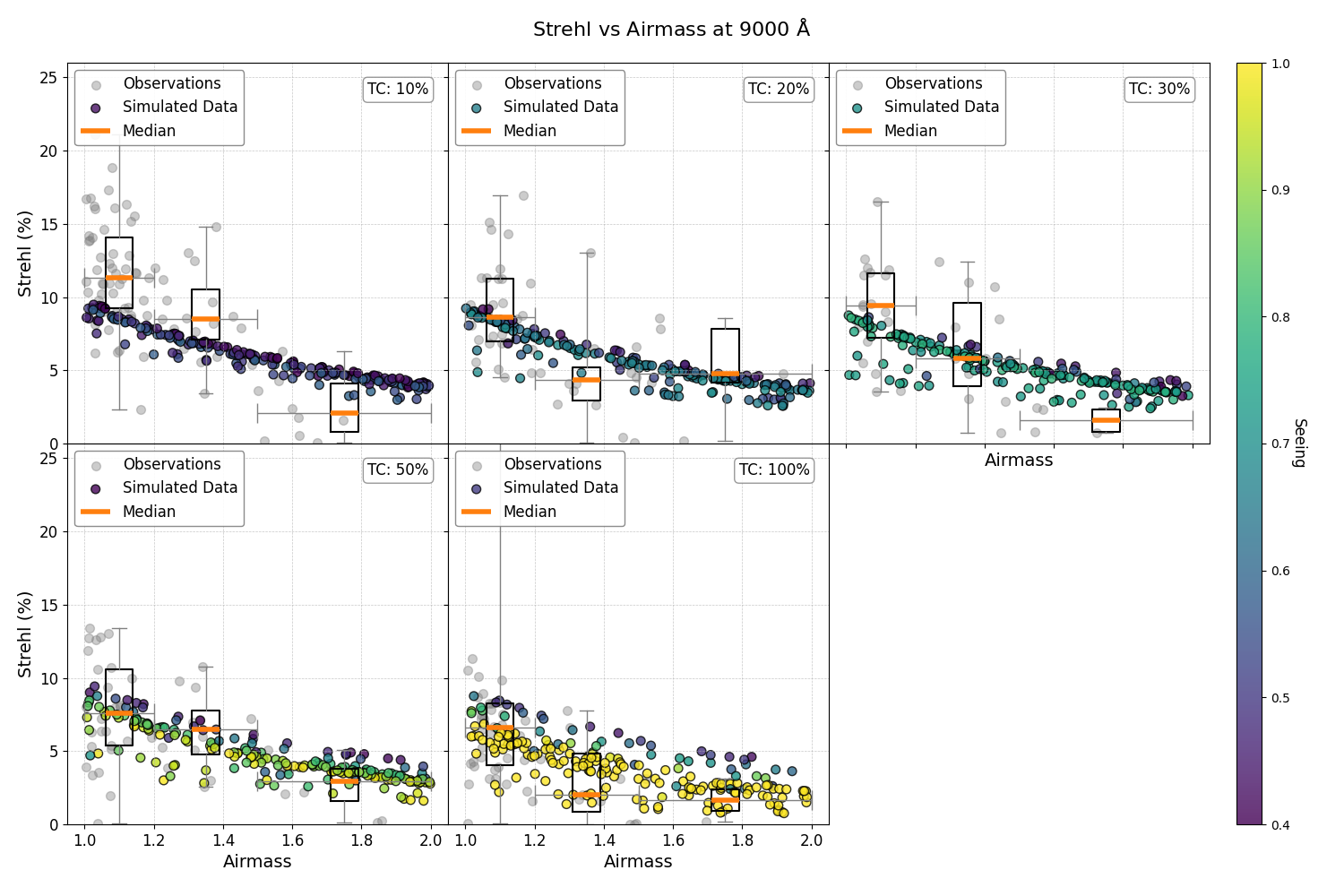}
    \caption{Observed and simulated Strehl ratios as a function of airmass and TC \qty{9000}{\angstrom}. Symbols as for Fig.~\ref{fig:FWHM700}}
    \label{fig:STR900}
\end{figure}

\begin{figure}
    \centering
    \includegraphics[width=0.85\textwidth]{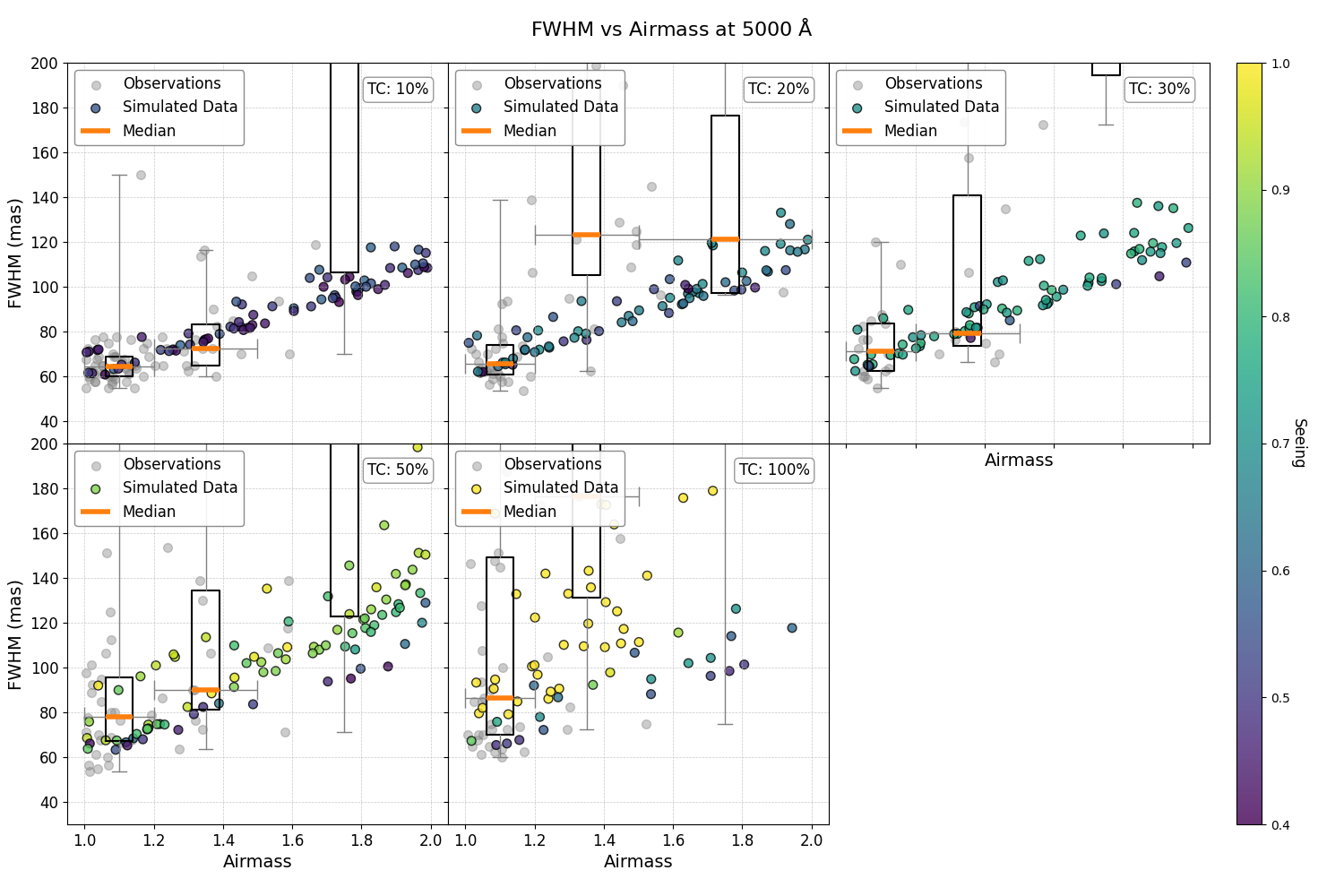}
    \caption{Observed and simulated FWHM as a function of airmass and TC at \qty{5000}{\angstrom} obtained using the new intercept described in Tab.~\ref{tab:ttblue}. Symbols as for Fig.~\ref{fig:FWHM700}}
    \label{fig:FWHM500_2}
\end{figure}

\begin{figure}
    \centering
    \includegraphics[width=0.85\textwidth]{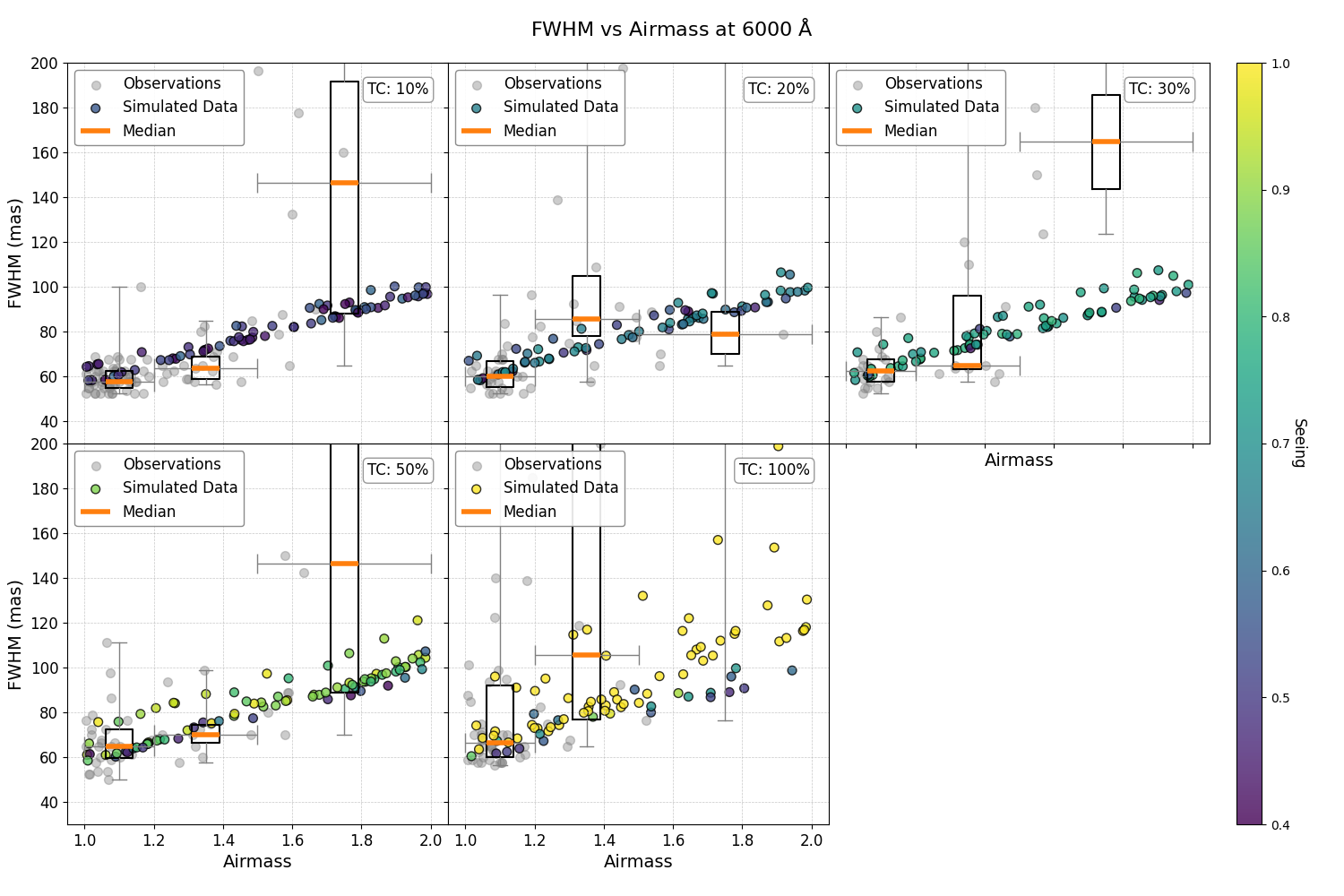}
    \caption{Observed and simulated FWHM as a function of airmass and TC at \qty{6000}{\angstrom} obtained using the new intercept described in Tab.~\ref{tab:ttblue}. Symbols as for Fig.~\ref{fig:FWHM700}}
    \label{fig:FWHM600_2}
\end{figure}

\end{appendix}

\end{document}